\documentclass[12pt]{article}
\usepackage{amssymb,amsmath}

\usepackage{curves}
\usepackage{epic}
\usepackage{eepic}
\usepackage{epsfig}

\newcommand{\dd}{\mbox{\rm d}}

\newcommand{\gam}{\gamma}

\newcommand{\DD}{\mbox{\rm D}}

\newcommand{\p}{\partial}

\newcommand{\be}{\begin{equation}}
\newcommand{\bear}{\begin{eqnarray}}

\newcommand{\ear}{\end{eqnarray}}
\newcommand{\ee}{\end{equation}}
\newcommand{\lbl}{\label}
\newcommand{\bi}{\bibitem}
\newcommand{\ci}{\cite}

\newcommand{\vs}{\vspace}

\begin{document}

\begin{center}

\
\vs{1cm}

\baselineskip .7cm

{\bf \Large Wheeler-DeWitt Equation in Five Dimensions\\ and Modified QED}

\vs{2mm}

\baselineskip .5cm
Matej Pav\v si\v c

Jo\v zef Stefan Institute, Jamova 39, SI-1000, Ljubljana, Slovenia; 

email: matej.pavsic@ijs.si

\vs{3mm}

{\bf Abstract}
\end{center}

\baselineskip .43cm
{\small

We consider the ADM splitting of the Einstein-Hilbert action in
five dimensions in the presence of matter that can be either a ``point
particle", or a set of scalar fields. The Hamiltonian, being a linear
superposition of constraints, is equal to zero. Upon quantization, we obtain
the Schr\"odinger equation for a wave functional, $\Psi$, that depends on the
matter degrees of freedom, and on the 5D gravity degrees of freedom.
After the Kaluza-Klein splitting, the functional
Schr\"odinger equation decomposes so that it contains a part due to 4D gravity,
a part due to electrodynamics, and a part due to matter. Depending on
choice of the matter term, we obtain two different versions of a
modified quantum electrodynamics. In one version, time automatically appears,
and there is no problem with infinite vacuum energy density of matter fields,
whereas in the other version such problems exist.
}

\baselineskip .55cm

\section{Introduction}

Quantization of gravity is still enigmatic. A straightforward approach
is to start from the Einstein-Hilbert action in the presence of matter.
Because of diffemorphism invariance, such system has constraints,
called the Hamilton and momentum constraints. In the quantized theory,
the constraints become operators that annihilate state vector. The
Hamilton constraint gives the Wheeler-DeWitt equations\,\ci{WheelerDeWitt}.
The Hamiltonian, $H$, which is a linear superposition of constraints
(this also involves the integration over space), is identically zero.
After quantization, the equation $H=0$ becomes $H|\Psi \rangle =0$, in which
there is no explicit time derivative term. How to obtain such a term is
subject of intensive research\,\ci{time}.

Another enigmatic subject is the unification of gravity with other fundamental
interactions. An approach that has been much investigated is to consider gravity
in a higher dimensional spacetime, $M_D$. The 4-dimensional gravity and
Yang-Mills interactions, including the electromagnetic U(1) interaction, are
all incorporated in the metric of $M_D$, if $M_D$ is equipped with appropriate
isometries\,\ci{Witten}.

As a first step, to see how the theory works, it is instructive to consider
gravity in five dimensions. Beciu\,\ci{Beciu}, Lacquantini and
Montani\,\ci{Montani} considered the
canonical gravity in 5D, by performing the ADM\,\ci{ADM} and Kaluza-Klein
splitting of spacetime. In this Letter we will extend their work to include a
matter term, $I_m$, in the action. Usually, a matter term consists of
scalar, $\varphi^\alpha$, or spinor fields, $\psi^\alpha$, minimally
coupled to gravity. Upon quantization, those fields and the conjugated
momenta become operators that create or annihilate particles.
In the Schr\"odinger representation, in which the field operators are
diagonal, the fields occur as arguments in the wave functional
$\Psi[\varphi^\alpha,...]$.

In a previous work\,\ci{PavsicWheeler}, we investigated an alternative approach.
The idea was based on the fact that, classically, objects are described by
spacetime coordinate functions $X^\mu$, $\mu=0,1,2,3$. The simplest object
is a point particle, described by $X^\mu (\tau)$. However, a point particle is
an idealization. In reality, there are no point particles. According
to Dirac\,\ci{DiracMembrane}, even the electron can be envisaged as a
charged spherical membrane, its center of mass being described by $X^\mu (\tau)$
(see\,\ci{BarutPavsic}). Neglecting the internal degrees of freedom, we
can describe a particle by an action functional $I_m [X^\mu (\tau)]$, bearing
in mind that such description is only valid outside the (extended) particle.
Because the particle is not a black hole, its radius is greater
than the Schwarzschild radius. Since the particle is
coupled to gravity, the total action contains the kinetic term for gravity,
$I_g [g_{\mu \nu}]$, as well. At the classical level, the degrees of
freedom are thus $X^\mu (\tau)$ and $g_{\mu \nu} (x^\rho)$. Extending
the theory to five dimensions, the classical degrees of freedom are
$X^M (\tau)$ and $G_{MN} (X^J)$, $M,N,J =0,1,2,3,5$. Such theory, besides
the constraints of the canonical gravity---now in 5D---has an additional
constraint, due to the representation invariance of the ``point particle"
action $I_m [X^M,G_{MN}]$. Upon quantization, the latter constraint becomes
the Klein-Gordon equation for a wave functional $\Psi [X^M,q_{ab}]$,
$a,b=1,2,3,5$, where instead of $G_{MN}$ we now consider the reduced number
of the metric components. We show how the Hamilton and momentum constraints,
if integrated over $\dd x^1 \dd x^2 \dd x^3 \dd x^5$ and split \`a la
Kaluza-Klein, contain quantum electrodynamics, appart from a difference
that comes from our usage of $I_m [X^M, G_{MN}]$, which leads to the terms
$-i \p \Psi/\p T$ and $-i \p \Psi/\p X^a$. The term $-i \p \Psi/\p T$ does
not necessarily give infinite vacuum energy.

We then also investigate the case in which the matter term is
$I_m [\varphi^\alpha,G_{MN}]$, $\alpha =1,2$. Upon quantization we have
constraints, acting on a state vector, and no time derivative term. But
otherwise, the constraints, integrated overs $\dd x^1 \dd x^2 \dd x^3 \dd x^5$,
closely match the Schr\"odinger representation of QED\,\ci{Hatfield},
appart from the term $H_g$ due to 4D gravity. We point out that, according to
the literature\,\ci{Kiefer}, the term $-i \p \Psi/\p T$ could come  from $H_g$
as an approximation.
So we obtain the Schr\"odinger equation for the evolution of a wave functional
that depends on the electromagnetic field potentials and scalar fields,
$\varphi^\alpha$. This is what we also have in the usual Schr\"odinger
(functional) representation\,\ci{Hatfield} of QED. 
Alternatively, we might
be interested in how evolves in time a wave functional that depends on
the 4D gravitational field and on the electromagnetic field. We show how
the time derivative term $-i \p \Psi/\p T$, i.e., the same term that
we obtain from $I[X^M,G_{MN}]$, results as an approximation to
the scalar field matter part, $H_m$, of the total Hamiltonian, $H$.
Regardles of which way we generate an approximative evolution term in the
quantum constraint equation, if matter consists of scalar (or spinors) fields,
then it gives infinite vacuum energy density coupled to gravity.

\section{ADM and Kaluza-Klein splitting of the Einstein-Hilbert action in the
presence of matter}

Let us consider the Einstein-Hilbert action in five dimensions in the presence
of a source, whose center of mass is described by $X^M (\tau)$, $M=0,1,2,3,5$:
\be
   I[X^A,G_{MN}]=M \int \dd \tau \, ({\dot X}^M {\dot X}^N G_{MN})^{1/2}
   +\frac{1}{16 \pi {\cal G}} \int \dd^5 x \, \sqrt{-G}\, R^{(5)} .
\lbl{2.1}
\ee 
Here $G_{MN}$ is the 5D metric tensor, $G$ its determinant, and ${\cal G}$
the gravitational constant in five dimension. The source is
not a point particle, it is an extended, ball-like or spherical membrane-like
object. We are not interested in the detailed dynamics of the coupling of
the ball or the membrane with the
gravitational field, we will only consider the center of mass. Therefore,
our description will be valid outside the object, whose radius may be small,
but greater than the Schwarzschild radius.

The metric tensor $G_{MN}$ can be split according to ADM \,\ci{ADM} as:
\be
   G_{MN} = \begin{pmatrix}  N^2-N^a N_a , & -N_a& \\
                    -N_b ,& - q_{ab} \\
                    \end{pmatrix} ,
\lbl{2.2}
\ee 
where $N=\sqrt{1/G^{00}}$ and $N_a=q_{ab} N^b=-G_{0a}$, $a=1,2,3,5$, are the laps and
shift functions in five dimensions.

Alternatively, $G_{MN}$ can be split according to Kaluza-Klein:
\be
 G_{MN} = \begin{pmatrix}  g_{\mu \nu}-\phi^2 A_\mu A_\nu , & k^2 \phi^2 A_\mu& \\
                  k^2 \phi^2 A_\nu,& - \phi^2 \\
                    \end{pmatrix} ,
\lbl{2.3}
\ee 
where $g_{\mu \nu}$ is the metric tensor, and $A_\mu$ the electromagnetic field
in 4D, whereas $k\equiv 2 \sqrt{{\cal G}^{(4)}}$ is a constant to be defined later.

From Eqs. (\ref{2.2}), (\ref{2.3}) we obtain the following relations:
\bear
    &&G_{00}=g_{00}-k^2 \phi^2 (A_0)^2 = N^2-N^a N_a \lbl{2.4}\\
    &&G_{0i}=g_{0i}-k^2 \phi^2 A_0 A_i = - N_i \lbl{2.5}\\
    &&G_{05} = k \phi^2 A_0 = -N_5  \lbl{2.6}\\
    &&G_{55} = -\phi^2 = - q_{55}   \lbl{2.7}\\
    &&G_{i5} = k \phi^2 A_i = -q_{i5}   \lbl{2.8}\\
    &&G_{ij} = g_{ij} - k^2 \phi^2 A_i A_j = - q_{ij}~,~~~~i,j =1,2,3 .   \lbl{2.9}
\ear
For the inverse metric tensors,
\be
   G^{MN} = \begin{pmatrix}  {1}/{N^2}, & -{N^a}/{N^2}& \\
                   -{N^b}/{N^2},& - {N^a N^b}/{N^2}-q^{ab} \\
                    \end{pmatrix} =
          \begin{pmatrix} g^{\mu \nu}, & k A^\mu& \\
                    k A^\nu,& k^2 A_\mu A^\mu - {1}/{\phi^2} \\
                    \end{pmatrix} ,
\lbl{2.10}
\ee     
we obtain
\bear
     &&G^{00} = g^{00} = \frac{1}{N^2} \lbl{2.11}\\
     &&G^{0i} = g^{0i} = - \frac{N^i}{N^2} \lbl{2.12}\\
     &&G^{05} =k A^0 = - \frac{N^5}{N^2}   \lbl{2.13}\\
     &&G^{55} = k^2 A_\mu A^\mu - \frac{1}{\phi^2}= \frac{(N^5)^2}{N^2}-q^{55}
     \lbl{2.14}\\
     &&G^{i5} = k A^i = - q^{i5}   \lbl{2.15}\\
     &&G^{ij} = g^{ij} - \frac{N^iN^j}{N^2} = q^{ij} . \lbl{2.16}
\ear
The 4D metric $g_{\mu \nu}$ can also be split according to ADM. This gives
the 3D metric, $\gam_{ij}$, and its inverse, $\gam^{ij}$.

The matter part of the action (\ref{2.1}) can be cast into the phase space
form,
\be
    I_m [X^M,p_M,\alpha,G_{MN}] = \int \dd \tau \, \left [p_M {\dot X}^M -
    \frac{\alpha}{2}(G^{MN} p_M p_N - M^2) \right ] ,
\lbl{2.17}
\ee
and split according to (\ref{2.2}),(\ref{2.10}). We obtain
\be
    I_m [X^M,p_M,\alpha,N,N^a,q_{ab}]= \int \dd \tau \left [p_M {\dot X}^M -
    \frac{\alpha}{2} \left (\frac{1}{N^2} (p_0 - N^a p_a)^2 - q^{ab} p_a p_b
    -M^2 \right ) \right ] .
\lbl{2.26}
\ee
Using the ADM splitting, the gravitational part of the action can be
written as
\be
    I_G[q_{ab},p^{ab},N,N^a] = \int \dd^5 x\,(p^{ab} {\dot q}_{ab} - 
    N {\cal H}_G - N^a {\cal H}_{G a} ).
\lbl{2.27}
\ee
Here
\bear
 &&{\cal H}_G = - \frac{1}{\kappa} Q_{abcd}p^{ab} p^{cd}
  + \kappa \sqrt{q} {\bar R}^{(4)} \lbl{2.28}\\
 &&{\cal H}_G^a = - 2D_b p^{ab}, \lbl{2.29}
\ear
where $\kappa = 1/(16 \pi {\cal G})$, and
 $Q_{abcd} =(1/\sqrt{q}) (- q_{ab} q_{cd}/(D-1) + q_{ac} q_{bd}+q_{ad} q_{bc})$
is the Wheeler-DeWitt metric in $D$-dimensions. In our case it is 
$D=q_{ab} q^{ab} = 4$.

Varying $I_G$ with respect to $p^{ab}$, we have the relation
\be
    p^{ab} = \kappa \sqrt{q}(K^{ab} - K q^{ab}),
\lbl{2.31}
\ee
where
\be
     K_{ab} = \frac{1}{2N} (-{\dot q}_{ab} +D_a^{(4)} N_b - D_b^{(4)} N_a).
\lbl{2.32}
\ee
Here,  ${\bar R}^{(4)}$ and $D_a^{(4)}$ are, respectively, the Ricci scalar
and the covariant derivative in the 4D space with the metric $q_{ab}$.

Our total phase space action
\be
     I = I_m + I_G
\lbl{2.33}
\ee
is a functional of the particle center of mass coordinates, $X^M$, of the
momenta, $p_M$, of the metric $q_{ab}$ on a 4D slice, of the momenta $p^{ab}$,
and of the set of the Lagrange multipliers, $\alpha$, $N$, $N^a$.
Variation of the total action with respect to $\alpha$, $N$ and $N^a$
gives the following constraints:
\bear
   &&\frac{1}{N^2}(p_0 - N^a p_a)^2 - q^{ab} p_a p_b -M^2 =G^{MN}p_M p_N -M^2=0,
\lbl{2.34}\\
  && -{\cal H}_G + \delta^3 ({\bf x} - {\bf X})\delta(x^5-X^5)
   \frac{1}{N} (p_0 - N^a p_a) = 0,
\lbl{2.35}\\
  && -{\cal H}_{G a} + \delta^3 ({\bf x} - {\bf X})\delta(x^5-X^5) p_a = 0.
\lbl{2.36}
\ear
In deriving the last two equation we have taken into account that
$(1/N^2)(p_0 - N^a p_a)= G^{0M} p_M = {\dot X}^0/\alpha$, and have
integrated the expressions
$$\int \dd \tau \, \frac{\alpha}{N^3}(p_0 - N^b p_b)^2 \delta^5 (x-X(\tau)),$$
and
$$\int \dd \tau \, \frac{\alpha}{N^2}p_a (p_0 - N^b p_b) \delta^5 (x-X(\tau)).$$
The integration $\int \dd^5 x \,\delta^5 (x-X(\tau)) =1$ was inserted into
$I_m$ in order to cast $I_m$ into a form, comparable to that of $I_G$.
Let me repeat that $X^M (\tau)$ are the center of mass coordinates of an
extended source, not of a point particle. The matter action (\ref{2.17}) is
thus an approximation to an action in which all other degrees of freedom of the
extended object have been neglected\footnote{See footnotes 1 and 2 of
ref.\ci{PavsicWheeler}}.

Eqs.\,(\ref{2.35}),(\ref{2.36}) are an infinite set of constraints, one 
at each point $x^a=({\bf x},x^5)\equiv {\bar x}$. If we multiply
Eqs.(\ref{2.35}),(\ref{2.36})
by ${\rm e}^{i k_a x^a}$, $a=1,2,3,5$, integrate over
$\dd^4 {\bar x}=\dd^3 {\bf x}\, \dd x^5$, we obtain the Fourier transformed
constraints, one for each $k_a$:
\bear
   &&-\int \dd^4 {\bar x}\, {\rm e}^{i k_b (x^b-X^b)} {\cal H}_G +\frac{1}{N}
   (p_0-N^b p_b)\bigl\vert_{X^a}=0,
\lbl{2.37}\\
 && -\int \dd^4 {\bar x}\, {\rm e}^{i k_b (x^b-X^b)} {\cal H}_{G a}
  +p_a\bigl\vert_{X^a}=0.
\lbl{2.38}
\ear

For $k_a=0$ (zero mode), and after fixing a gauge $N=1,~N^a=0$,
 Eqs.\,(\ref{2.37}),(\ref{2.38}) become
\bear
   &&\int \dd^4 {\bar x}\,  {\cal H}_G = p_0,
\lbl{2.39}\\
  && \int \dd^4 {\bar x}\,  {\cal H}_{G a} = p_a.
\lbl{2.39a}
\ear
Using (\ref{2.28}),(\ref{2.29}), we have
\bear
  && - \frac{1}{\kappa} \int \dd^4 {\bar x} \, (Q_{abcd}\, p^{ab} p^{cd} 
  + \kappa \sqrt{q} {\bar R}^{(4)}) = p_a ,  \lbl{2.40}\\
   && - 2 \int \dd^4 {\bar x} \, D_b {p_a}^b 
   = -2 \oint \dd \Sigma_b {p_a}^b = p_a .
\lbl{2.41}
\ear

Splitting the above equations \`a la Kaluza-Klein by using
Eqs.\,(\ref{2.4})--(\ref{2.16}), it turns out that they contain the parts
of the 4D gravity and the Maxwell theory. Eq.\,(\ref{2.40}) can be written as
\be
   H_G= \int \dd^3 x\,({\cal H}_g + {\cal H}_{EM} + {\cal H}_{\phi})= p_0
\lbl{2.42}
\ee
where according to Ref.\,\ci{Montani}
\bear
     &&{\cal H}_g = - \frac{1}{\kappa^{(4)}} T_{ijk \ell} \pi^{ij} \pi^{k \ell}
     + \kappa^{(4)} \sqrt{\gamma} R^{(3)} , \lbl{2.43}\\
     &&{\cal H}_{EM} = - \frac{2}{\kappa^{(4)} \sqrt{\gam} k^2 \phi^3}
     \pi^i \pi^j \gam_{ij} -\frac{\kappa^{(4)}}{4} \sqrt{\gam} k^2 \phi^3
     F_{ij} F^{ij}  \lbl{2.44}\\
     &&{\cal H}_\phi = - 2 \kappa^{(4)} \sqrt{\gamma} \DD^i \DD_i \phi -
     \frac{1}{6\kappa^{(4)} \sqrt{\gamma}} \pi_\phi^2 +
     \frac{1}{3\kappa^{(4)} \sqrt{\gamma}} \pi_\phi \pi^{ij} \gam_{ij} ,
\ear
with
$T_{ijk \ell}=(\gam_{ik}\gam_{j \ell} + \gam_{i \ell}
\gam_{jk} - \frac{2}{3}\gam_{ij} \gam_{k \ell})$, $i,j,k, \ell=1,2,3$,
whereas $\pi^{ij}$, $\pi^i$, and $\pi_\phi$ are the
canonical momenta conjugated to the spatial metric $\gam_{ij}$, the
electromagnetic potential $A_i$, and the scalar field $\phi$, respectively.

Eq.\,(\ref{2.41}) can be split according to
\be
     -2 \int \dd^4 {\bar x} (D_i {p_a}^i + D_5 {p_a}^5) = p_a
\lbl{ 2.45}
\ee
Let us assume that $D_5 {p_a}^5 = 0$, because of  the isometry along the 5th
dimension (cylindricity condition). Then, for $a=j$, we have
\be
  -2 \int \dd^4 {\bar x} D_i {p_j}^i = -2 \oint \dd \Sigma_i \, {p^i}_j = p_j
\lbl{2.46}
\ee
where ${p^i}_j$ can be split into the part due to the spatial metric
$\gam^{ij}$, the part due to the electromagnetic field $A_i$, and the part
due to the scalar field $\phi$ (see Ref.\, \ci{Montani}).

For $a=5$, using (\ref{2.31}),(\ref{2.32}), we find:
\bear
     -2 \oint \dd \Sigma_i \, {p_5}^i &=& \oint \dd \Sigma_i \kappa \sqrt{q}
     \left [ -\gam^{ij} \frac{\dd}{\dd t} (k \phi^2 A_j) + k A^i 
     \frac{\dd}{\dd t} (\phi^2) \right ] \nonumber\\
     &=& - \oint \kappa^{(4)} k \phi^3 \sqrt{\gam} \, \dd S_i {\dot A}^i  = p_5.
\lbl{2.47}
\ear
Here the hypersurface element in 4-space has been factorized
according to $\dd \Sigma_i = \dd S_i \dd x^5$, and the determinant
according to $\sqrt{q}= \phi \sqrt {\gam}$. The integration over $\dd x^5$
then leaded to $\int \kappa \dd x^5 \equiv \kappa^{(4)} \equiv 
\int \dd x^5/(16 \pi {\cal G}) \equiv 1/(16 \pi {\cal G}^{(4)})$.

Bear in mind that we have chosen the gauge $N=1$, $N^a=0$, which also
implies $N_a = q_{ab} N^b =0$. Then, from Eq.\,(\ref{2.6}) it follows
$A_0 = 0$ This is the temporal gauge for the electromagnetic potential.
Therefore, the electromagnetic field, $F_{\mu \nu} = \p_\mu A_\nu -
\p_\nu A_\mu$, has the components $F_{0i} = \p_0 A_i - \p_i A_0$
$=\p_0 A_i \equiv {\dot A}_i = E_i$. Eq.\,(\ref{2.47}) then reads
\be
   - \oint  \kappa^{(4)} k \phi^3 \sqrt{\gam} \, \dd S_i E^i = p_5.
\lbl{2.48}
\ee
Because in the Kaluza-Klein theory the 5th component of a particle's momentum
is the electric charge, Eq.\,(\ref{2.48}) is the Gauss law of electrodynamics.

\section{Quantization}

After quantization, the classical constraints (\ref{2.34})--(\ref{2.36})
become the conditions on a state $|\Psi \rangle$:
\bear
    &&(- G^{MN} p_M p_N - M^2)|\Psi \rangle = 0 , \lbl{3.1}\\
    &&\frac{1}{\kappa} (Q_{abcd} p^{ab} p^{cd} - \kappa \sqrt{q} 
    {\bar R}^{(4)} |\Psi \rangle = - \delta^4 ({\bar x}
     - {\bar X}) p_0 |\Psi \rangle, \lbl{3.2}\\
     && - 2 q_{a c} \DD_b p^{cb} |\Psi \rangle = \delta^4 ({\bar x}-{\bar X})
     p_a |\Psi \rangle, \lbl{3.3}
\ear
where $p_M$, $p^{ab}$ are now momentum operators, and
$\delta^4 ({\bar x}-{\bar X}) \equiv
\delta^3 ({\bf x}-{\bf X}) \delta (x^5-X^5)$, ${\bar x}\equiv x^a$, ${\bar X}
\equiv X^a$, $a=1,2,3,5$. The state $|\Psi \rangle$ can be represented as
a wave function(al) $\langle T,X^a,q_{ab}|\Psi \rangle$ 
$\equiv \Psi[T,X^a,q_{ab}]$,
and the momentum operators as $p_M = -i \p/\p X^M$,
$p^{ab}=-i\delta/\delta q_{ab}$. Integrating (\ref{3.2}) and (\ref{3.3})
over $\dd^4 {\bar x} \equiv \dd^3 {\bf x}\, \dd x^5$ gives\footnote{Here we
neglect the ordering ambiguity issues.}
\bear
  &&\frac{1}{\kappa} \int  \dd^4 {\bar x} \left ( 
  - Q_{abcd} \frac{\delta^2}{\delta q_{ab} \delta q_{cd}}
  - \kappa \sqrt{q} {\bar R}^{(4)} \right ) \Psi = i \frac{\p \Psi}{\p T} ,
  \lbl{3.4}\\
  && - 2 \int \dd^4 {\bar x} \, q_{cb} \DD_b 
  \left ( -i \frac{\delta \Psi}{\delta q_{cb}} \right )  
  = - i \frac{\p \Psi}{\p X^a} .  \lbl{3.5a}
\lbl{3.5}
\ear
Every solution to the quantum constraints (\ref{3.1})--(\ref{3.3}) satisfies
the Schr\"odinger equation (\ref{3.4}) with the time $T\equiv X^0$.
The opposite is not true: not every solution of the Schr\"odinger equation
(\ref{3.4}) does satisfy the full set of constraints (\ref{3.1})--(\ref{3.3}).
There is no term that could give
infinite energy coupled to the 5D gravity. Instead of such annoying term, we
have the term $i \p \Psi/\p T$.

We can envisage that there exists a particular, wave packet-like
solution, $\Psi [T,X^a,q_{ab}]$,
that describes a 5D spacetime, split \`a la Kaluza-Klein. Then
Eqs.\,(\ref{3.1})--(\ref{3.5}) contain the pieces that correspond to the 4D gravity,
to the electromagnetic field, and to the scalar field $\phi \equiv -G_{55}$.
For instance, Eq.\,(\ref{3.4}) can then be written in the form
\be
    H\left (-i \frac{\delta}{\delta \gam_{ij}},-i \frac{\delta}{\delta A_i}
     -i\frac{\delta}{\delta \phi}
    \right ) \Psi[T,X^i,\gam_{ij}, A_i, \phi] =
    i \frac{\p}{\p T}\Psi[T,X^i,\gam_{ij}, A_i, \phi] .
\lbl{3.6}
\ee
The fifth component of Eq.\,(\ref{3.5}) then becomes
\be
   - \int \dd^3 {\bf x}\, \phi^3 \sqrt{\gam} \,
   \p_i \left ( -i \frac{\delta \Psi}{\delta A_i} \right )
   = - i \frac{\p \Psi}{\p X^5} = e \Psi.
\lbl{3.7}
\ee
The above equations are the quantum versions of the classical equations
(\ref{2.42})--(\ref{2.48}).

In addition, the state $|\Psi \rangle $ also satisfies Eq.\,(\ref{3.1}), i.e.,
the 5D Klein-Gordon equation
\be
    (- G^{MN} {\cal D}_M {\cal D}_N - M^2 )\Psi = 0,
\lbl{3.8}
\ee
that, after the Kaluza-Klein splitting becomes
\be
  \left [ g^{\mu \nu} (-i {\cal D}_\mu^{(4)} + e A_\mu) (-i {\cal D}_\nu^{(4)} 
  + e A_\nu) - m^2 \right ] \Psi = 0 ,
 \lbl{3.9}
\ee
where $m^2=M^2 + e^2/\phi^2$, and ${\cal D}_\mu^{(4)}$ the covariant derivative
with respect to the 4D metric $g_{\mu \nu}$.

Eq.\,(\ref{3.6}) generalizes the functional Schr\"odinger equation
for the electromagnetic field\,\ci{Hatfield}, whereas Eq.\,(\ref{3.7})
generalizes the Gauss law constraint. 

\section{Arbitrary matter term in the action}

In general, the matter term, $I_m$, of the action is a functional of
a set of fields $\varphi^\alpha$. So we have the following total action:
\be
   I = I_G [G^{MN}] + I_m [\varphi^\alpha, G^{MN}]
\lbl{4.1}
\ee
For instance, if $\alpha=1,2$, then $\varphi^\alpha$ can be the real
an imaginary component of the charged scalar field. The matter action is
then
\be
   I_m = \mbox{$\frac{1}{2}$} \int \dd^5 x \, \sqrt{-G} (G^{MN}
          \p_M \varphi^\alpha \p_N \varphi_\alpha - M^2 \varphi^\alpha
          \varphi_\alpha ).
\lbl{4.2}
\ee
After the ADM splitting, we have
\be
     I_m = \mbox{$\frac{1}{2}$} \int \dd t \, \dd^4 {\bar x} \,
          N \sqrt{q} \left [ \left ( \frac{1}{N} \right )^2
          ({\dot \varphi}^\alpha - N^a \p_a \varphi^\alpha)        
          ({\dot \varphi}_\alpha - N^b \p_b \varphi_\alpha)
          - q^{ab} \p_a \varphi^\alpha \p_b \varphi_\alpha 
          - M^2 \varphi^\alpha \varphi_\alpha \right ] .
\lbl{4.3}
\ee

The Hamiltonian, corresponding to the action (\ref{4.1}) is
\be
   H = - \int \dd^4 {\bar x} \left ( N \frac{\delta I}{\delta N}
     +N^a \frac{\delta I}{\delta N^a} \right ) ,
\lbl{4.4}
\ee
where  $-\delta I/\delta N = {\cal H} = {\cal H}_G +{\cal H}_m$, and
$-\delta I/\delta N^a = {\cal H}_a = {\cal H}_{G\, a} +{\cal H}_{m\, a}$ are
the constraints.

Here $H_m = \int \dd^4 {\bar x}\, {\cal H}_m$ is the Hamiltonian for the matter
fields. In the case in which $I_m$ is given by Eq.\,(\ref{4.3}), it is
\be
   H_m = -\int \dd^4 {\bar x} \, \frac{\delta I_m}{\delta N}
      =  \mbox{$\frac{1}{2}$} \int \, \dd^4 {\bar x} \,
      \frac{1}{\sqrt{q}} (\Pi^\alpha \Pi_\alpha 
      + q^{ab}\p_a \varphi^\alpha \p_b \varphi_\alpha
      + M^2 \varphi^\alpha \varphi_\alpha ) ,
\lbl{4.6}
\ee
where
\be
    \Pi_\alpha = \frac{\p {\cal L}_m}{\p {\dot \varphi}^\alpha}
      = \frac{\sqrt{q}}{N} ({\dot \varphi}_\alpha - N^a \p_a \varphi_\alpha ).
\lbl{4.7}
\ee

Upon quantization, we have
\be
    (H_G + H_m) |\Psi \rangle = 0 .
\lbl{4.8}
\ee
In the usual approaches to quantum field theories, where gravity is not taken
into account, one does not assume the validity of the constraint equation
Eq.\,(\ref{4.8}), but of the Schr\"odinger equation
\be
    H_m |\Psi \rangle = i \frac{\p |\Psi \rangle}{\p t} .
\lbl{4.9}
\ee
But we see, that within the more general setup with gravity, the validity
of the Schr\"odinger equation (\ref{4.9}) cannot be taken for granted.
Eq.\,(\ref{4.9}) is presumably incorporated in the constraint equation
(\ref{4.8}), and this has to be derived. Various authors have worked
on such problem\,\ci{Kiefer} of how to derive $i \p |\Psi \rangle/\p T$
from $H_G$.

The opposite, namely how to derive $i \p |\Psi \rangle/\p T$ from
$H_m$ in order to obtain from (\ref{4.8}) the equation
$H_G = i \p |\Psi \rangle/\p T$, is also an interesting problem. There is
a lot of discussion in the literature on such problem\ci{Rovelli}. Let me
show here a possible procedure.
Despite that our procedure refers to the 5D gravity, it holds also
for the usual, 4D, gravity.

From the stress-energy tensor
\be
    T^{MN} = a \left [ \p^M \varphi^* \p^N \varphi - \mbox{$\frac{1}{2}$}
    G^{MN} (G^{JK} \p_J \p_K -M^2 \varphi^* \varphi) \right ] ,
\lbl{4.10}
\ee
after taking the Ansatz
\be
\varphi = A \, {\rm e}^{i S} ,
\lbl{4.11}
\ee
we obtain the
following expression for the field momentum:
\be
   P^M = \int \sqrt{-G}\, \dd \Sigma_N  \, T^{MN} 
   = a \int \sqrt{-G} \,\dd \Sigma_N \, A^2 \, \p^M S \, \p^N S  .
\lbl{4.12}
\ee
Here we have taken into account that $\varphi$ satisfies the Klein-Gordon
equation, which in the limit $\hbar \rightarrow 0$
gives $\p_M S\, \p^M S - M^2 =0$, implying that the second
term in Eq.\,(\ref{4.10}) vanishes.

Let us now assume that $|\varphi|=A^2$ is picked around the classical
particle worldline. As a convenient approximation let us take
\be
    A^2 = \int \dd \tau \, \frac{\delta^5 (x-X(\tau))}{\sqrt{_G}} .
\lbl{4.13}
\ee

Since $p_N=\p_N S$, we obtain
\be
   P^M = a \int \dd \Sigma_N \, \dd \tau \, \delta^5 (x-X(\tau)) p^M p^N .
\lbl{4.14}
\ee
Assuming that $\dd \Sigma_N = p_N/\sqrt{p^2} \dd \Sigma$, where
$\dd \Sigma = \dd^4 {\bar x}$, taking a gauge $X^0 = \tau$, i.e.,
${\dot X}^0=1$, and integrating over $\tau$, we find
\bear
   P^M &=& a \int \dd \Sigma \frac{p_N p^N}{\sqrt{p^2}}\,p^M \,
   \frac{\delta^4 ({\bar x}-{\bar X})}{|{\dot X}^0|} \nonumber \\
   &=& a \int \dd^4 {\bar x} \, M p^M \delta^4 ({\bar x}-{\bar X})
   = a M p^M = p^M .
\lbl{4.15}
\ear
We see that the field momentum is equal to the particle's momentum, if the
normalization constant is $a=1/M$.

Alternatively, if we do not integrate over $\tau$ in Eq.\,(\ref{4.14}),
we have
\be
    P^M = a \int \dd \Sigma \, \dd \tau\, M \, p^M \delta^5 (x-X(\tau)) .
\lbl{4.16}
\ee
In the gauge in which $\tau=x^0$, it is $\dd \Sigma \, \dd \tau=
\dd^4 {\bar x} \dd x^0 = \dd^5 x$. Integrating over $\dd^5 x$,
we obtain the same result as in Eq.\,(\ref{4.15}).

This was a classical theory. Upon quantization, the momentum becomes the
operator $p_M=-i \p/\p X^M$,
in particular, $p_0=-i \p/\p X^0 \equiv -i \p/\p T$. Then Eq.\,(\ref{4.8})
becomes
\be
    (H_G - i \frac{\p}{\p T}|\Psi \rangle = 0 ,
\lbl{4.18}
\ee
which corresponds to our equation (\ref{3.4}), derived from the total
action (\ref{2.33}) with the point particle matter term.

Since we consider a five or higher dimensional spacetime, we can perform
the Kaluza-Klein splitting. Then Eq.\,(\ref{4.8}) contains the terms due to the
4D gravity and the terms due to the electromagnetic or Yang-Mill fields:
\be
   (H_g + H_{EM} + H_m + .... ) |\Psi \rangle = 0.
\lbl{4.19}
\ee
All those terms together form a constraint on a state vector. There is no
explicit time derivative term. We have two basically different possibilities:

(a) A time derivative term comes from $H_g$ as an approximations.
Then the system (\ref{4.19}) becomes the Schr\"odinger equation for
the electromagnetic field in the presence of ``matter":
\be
   \left ( -i \frac{\p}{\p T} + H_{EM} + H_m \right ) |\Psi \rangle = 0.
\lbl{4.19a}
\ee
We have considered the case in which matter consists of a charged scalar field.
We could as well consider a spinor field.

(b) A time derivative term comes from $H_m$ as an approximation. Then
Eq.\, (\ref{4.19}) describes the evolution of the electromagnetic and
the gravitational field:
\be
   \left ( H_g +H_{EM} - i \frac{\p}{\p T} \right ) |\Psi \rangle = 0.
\lbl{4.20}
\ee

In general, both equations, (\ref{4.19a}) and (\ref{4.20}) are
approximations to the constraint (\ref{4.19}). In particular, if for the matter
term in the classical action (\ref{4.1}), instead of 
$I_m [\varphi^\alpha, G^{MN}]$, we take the ``point particle" action
$I_m [X^M, G^{MN}]$, then---as shown in Secs.\,2 and 3---we also arrive
at Eq.\,(\ref{4.20}). This is then an ``exact" equation, because
the term $-i \p/\p T$ comes directly from $p_0$ of the ``point particle".

If in Eq.\,(\ref{4.19}) we do not ticker with the term $H_m$,
but leave it as it is, then it gives infinite vacuum energy.

\section{Discussion}

We have considered five dimensional gravity in the presence of a source
whose center of mass was described by a point particle action. After
performing the ADM splitting and varying the action with respect to the
lapse and shift functions, we obtained the Hamiltonian constraint and
four momentum constraints. In addition, we also obtained the constraint
coming from the reparametrization invariance of the point particle term
in the total action. In the quantized version of the theory, all those
constraints act on a state that can be represented as $\Psi[T,X^a,q_{ab}]$,
a function(al) of the particle's coordinates $X^M=(T,X^a)$, and of
of the 4D metric, $q_{ab}$, $a,b=1,2,3,5$. The $\Psi$ satisfies
the Wheeler-DeWit equation in which the term due to the presence of the
particle is $-i \p \Psi/\p T$. It also satisfies quantum momentum
constraints with a term $-i \p/\p X^a$. Besides that, the
$\Psi[X^M,q_{ab}]$ satisfies the Klein-Gordon equation in curved space.
Also in the usual theories the Klein-Gordon field in a curved space
is a functional of the (background) metric. In our approach the metric is
not a background metric. It is a dynamical metric, therefore the wave
functional $\Psi[X^M,q_{ab}]$ satisfies the Wheeler-DeWitt equation
as well.

If we split the 5D metric \`a la Kaluza-Klein, then the equations
split into the terms describing the 4D gravity and electrodynamics.
In the quantized theory we obtain the functional representation of
quantum electrodynamics in the presence of gravity. But there are some
subtleties here, because
according to the usual theory\,\ci{Hatfield}, also a term due to the
stress-energy of a charged scalar field or a spinor field should be present
in Eq.\,(\ref{3.6}).
There is no such term in Eq.\,(\ref{3.6}), because we have started from the
classical action (\ref{2.1}) with a ``point particle" matter term. The
corresponding stress-energy tensor has---amongst others---the five components
$T_{00}$, $T_{0a}$, $a=1,2,3,5$, as given in Eqs.\,(\ref{2.35}),(\ref{2.36}).
Integrating over $\dd^4 {\bar x}$, we obtain the particle's 5-momentum
$(p_0,p_a)$ that, after quantization becomes $(-i \p/\p T, -i \p/\p X^a)$.
The term $i \p/\p T$ in the Schr\"odinger equation (\ref{3.2}) thus comes
from the stress-energy of a ``point particle".

In the usual approaches, one does not start from the action (\ref{2.1})
with a ``point particle" matter term, but from an action with a charged
scalar field, $\varphi$, or a spinor field, $\psi$. In Sec.\,4 we
explored how this works in five dimensions.  The Kaluza-Klein
splitting of the 5D gravity in the presence of a charged scalar or spinor
field gives, after quantization, a wave functional equation (\ref{4.19})
without the
time derivative term. In such approach, the notorious ``problem of time"
remains\footnote{Moreover, because of the infinite vacuum energy density
of the charged scalar or spinor field coupled to gravity, there is the
problem of the cosmological constant.}. On the other hand, in the textbook
formulation\,\ci{Hatfield} of the Schr\"odinger representation of quantum
electrodynamics that is not derived from a 5D or a higher dimensional gravity,
one has the term $i \p \Psi/\p T$, besides the energy term due to
$\varphi$ or $\psi$. According to the
existing literature\,\ci{Kiefer}, such time derivative term can
occur from the gravitational part of the total Hamiltonian. So we obtained
Eq.\,(\ref{4.19a}).
We have also shown how the matter part of the Lagrangian with
the scalar fields can give the time derivative term. Thus we obtained
Eq.\,(\ref{4.20}). So we have a relation between the approach that
starts from the classical action $I[X^M, G_{MN}]$, and the usual approach
that start form $I[\varphi^\alpha,G_{MN}]$. But there is a crutial difference,
because in the former approach, after quantization, a wave functional
$\Psi[X^M,q_{ab}]$ satisfies the Klein-Gordon equation and the
Wheeler-DeWitt equation, whereas in the latter approach we have a wave
functional $\Psi[\varphi^\alpha,q_{ab}]$ that satisfies the Wheeler-DeWitt
equation only.

Having in mind that we usually consider a classical theory and its
quantization, it seems natural to start from classical objects, e.g.,
particles, described by $X^M$, coupled to gravity, described by $G_{MN}$,
so that after quantization we obtain a wave functional $\Psi[X^M,q_{ab}]$.
Having a wave functional $\Psi[X^M,q_{ab}]$, we can envisage its second
quantization, so that $\Psi$ and its Hermitian conjugate are related to
the operators that create at $X^M$ a particle with a surrounding gravitational
field $q_{ab}$. This brings new directions for further development
of quantum field theories, including gravitational, electromagnetic, and
Yang-Mills fields that arise in higher dimensional spacetimes.

\section{Conclusion}

From the Wheeler-DeWitt equation in five dimensions we have obtained,
depending on choice of a matter term, two different
versions of modified quantum electrodynamics in the Schr\"odinger
representation. 
The five dimensional gravity with matter was
only a toy model. A more realistic theory, describing all fundamental
interactions, should be formulated in higher dimensions\,\ci{Witten}. Since
QED is a theory that in many respects works very well, this indicates that
also the higher dimensional Wheeler-DeWitt equation, into which QED is
embedded, could be---to a certain extent--- a valid
description of Nature. On the other hand, for many reasons
gravity---regardless of the spacetime dimensionality---cannot be considered
as a complete, but rather as an effective theory arising from a more fundamental theory. The underlying
more fundamental theory could have roots in any of the currently
investigated fields of research such as strings\,\ci{strings},
branes\,\ci{Duff}, brane worlds\,\ci{TapiaPavsic},
loop quantum gravity\,\ci{LoopQG},
gravity as entropic force\,\ci{Verlinde},
etc. There could also
be some new, not yet explored landscape of theoretical
physics\,\ci{PavsicBook}--\ci{PavsicSymplectic}.

\vs{4mm}

\centerline{\bf Acknowledgment}

This work has been supported by the Slovenian Research Agency.

\end{document}